\documentclass[12pt]{article}
\usepackage{graphicx}
\usepackage{graphics}
\usepackage{amsthm}
\usepackage{amssymb, amsmath}
\title{Generalized raising and lowering operators for supersymmetric quantum mechanics}
\author{Mark W. Coffey\\
Department of Physics\\
Colorado School of Mines\\
Golden, CO  80401\\
USA\\
mcoffey@mines.edu}
\date{January 9, 2015} 
\begin{document}
\maketitle
\baselineskip=25 pt
\begin{abstract}

Supersymmetric quantum mechanics has many applications, and typically uses a raising and
lowering operator formalism.  For one dimensional problems, we show how such raising and lowering operators may be generalized to include an arbitrary function.  As a result, the usual Rodrigues'
formula of the theory of orthogonal polynomials may be recovered in special cases, and it may
otherwise be generalized to incorporate an arbitrary function.  We provide example generalized
operators for several important classical orthogonal polynomials, including Chebyshev, 
Gegenbauer, and other polynomials.  In particular, as concerns Legendre polynomials and associated Legendre functions, we supplement and generalize results of Bazeia and Das.

\end{abstract}
 
\vspace{.25cm}
\baselineskip=15pt
\centerline{\bf Key words and phrases}
\medskip 
raising operator, lowering operator, supersymmetry, Rodrigues' formula, Legendre polynomial,
Laguerre polynomial, Chebyshev polynomial, Gegenbauer polynomial

\bigskip
\bigskip
\noindent
{\bf 2010 PACS codes}
\newline{02.30.Gp, 03.65.-w, 02.30.Hq} 

\baselineskip=25pt

\pagebreak
\centerline{\bf Introduction} 

\bigskip

The concepts of supersymmetry from quantum field theory have led to diverse applications
in many other areas of physics including statistical, condensed matter, and nuclear physics, and
quantum mechanics \cite{cks,gend,junker}.  The supersymmetric scheme permits to combine fermionic and bosonic degrees of freedom.  Supersymmetry presents a hierarchy of Hamiltonians, and for 
nonrelativistic quantum mechanics provides factorizations of the Schr\"{o}dinger differential
equation.  Factorization of the Hamiltonian is done with operators incorporating the 
superpotential $W$, and $W$ satisfies Ricatti equations with the partner potentials being 
the inhomogeneous term.

The factorization of Hamiltonians is typically developed via raising and lowering operators,
each being of first order.  In this paper we focus on the one dimensional Schr\"{o}dinger equation
and describe how an arbitrary function, denoted $h(x)$, may be incorporated into these operators.  This procedure allows known Rodrigues' formulae to be recovered in short order for even the 
simplest choices of $h(x)$, and otherwise gives a way to develop generalized Rodrigues' formulae 
and other identities.  In this Introduction we provide illustrations for the quantum simple 
harmonic oscillator, and the associated Hermite and Laguerre polynomials.  We then concentrate
on the Legendre differential equation, thereby generalizing many of the results of \cite{bdas}.
We also supplement \cite{bdas} by giving proofs of certain identities.  Furthermore, we provide
a generalization to Gegenbauer polynomials with parameter $\lambda$.
Our method is applicable in a great many contexts, and we briefly mention other examples at the
end.

We let $L_n^\alpha(x)$ denote the $n$th Laguerre polynomial, $H_n(x)$ the $n$th Hermite polynomial,
$(z)_n=z(z+1)\cdots(z+n-1)=\Gamma(z+n)/\Gamma(z)$ the Pochhammer symbol, with $\Gamma$ the Gamma function (e.g., \cite{nbs,andrews,grad,rainville}).  
We recall that Hermite polynomials are special cases of Laguerre polynomials, such that
$$H_{2n}(x)=(-1)^n 2^{2n}n!L_n^{-1/2}(x^2) ~~\mbox{and}~~ H_{2n+1}(x)=(-1)^n 2^{2n+1}n!xL_n^{1/2}(x^2) \eqno(1.1)$$
(e.g., \cite{grad}, p. 1037).  Both of these types of polynomials are instances of the
confluent hypergeometric function $_1F_1$.

We work with dimensionless quantities throughout, so that for the quantum harmonic oscillator
the raising and lowering operators may be taken as $a^\pm =x \mp {d \over {dx}}$.
For the raising operator, we make the ansatz 
$$a^+=x-{d \over {dx}}=f_1 {d \over {dx}}g_2+h,$$
wherein the functions $f_1$ and $g_2$ are to be determined in terms of the arbitrary 
function $h$.  Letting $a^+$ act on a differentiable function $f$, and then equating coefficients 
of $f$ and $f'$, we find that, aside from multiplicative constants,
$$g_2(x)=\exp\left[-\int(x-h)dx\right],$$
and
$$f_1(x)=-{1 \over {g_2(x)}}=-\exp\left[\int(x-h)dx\right].$$
It goes very similarly for generalizing $a^-$, so that with a sign change the roles of
$f_1$ and $g_2$ are reversed.  
The original operators $a^\pm$ do not depend upon the energy level $n$ and accordingly neither 
do the generalized operators with $h$ in general different from $x$.  The well known equal
energy spacings follow.

As usual, the ground state obtains from $a\psi_0=0$ and $\psi_0(x) \propto \exp(-x^2/2)$.
Acting successively on the ground state wavefunction with the raising operator provides the Hermite functions.  If $h=0$, then $a^+=-e^{x^2/2}{d \over {dx}}e^{-x^2/2}$.
By multiplying by $e^{x^2/2}$ and imposing a sign convention, even in the simplest case of
taking $h=0$, the usual Rodrigues' formula for Hermite polynomials is recovered.

For the Laguerre differential equation
$$xu''+(\alpha-x+1)u'+nu=0, \eqno(1.2)$$
raising and lowering operators may be taken as
$$A_+=x{d \over {dx}}-x+\alpha+n, ~~~~A_-=-x{d \over {dx}}+n, \eqno(1.3)$$
so that the differential equation assumes the forms
$$A_+A_-L_n^\alpha(x)=n(n+\alpha)L_n^\alpha(x)$$
and
$$A_-A_+L_{n-1}^\alpha(x)=n(n+\alpha)L_{n-1}^\alpha(x).$$
Here we have applied recurrence relations (\cite{grad}, 8.971.3) so that
$A_+L_{n-1}^\alpha(x)=nL_n^\alpha(x)$ and $A_-L_n^\alpha(x)=(n+\alpha)L_{n-1}^\alpha(x)$.
If we put $A_+=f_1 {d \over {dx}}g_2+h$, then we find that, omitting a multiplicative constant,
$$g_2(x)=x^{\alpha+n}e^{-x}\exp\left[-\int {{h(x)} \over x}dx\right],$$
and
$$f_1(x)={x \over {g_2(x)}}=x^{1-\alpha-n}e^x\exp\left[\int {{h(x)} \over x}dx\right].$$
The correspondence to the usual Rodrigues' formula 
$L_n^\alpha(x)={1 \over {n!}}e^x x^{-\alpha} \left({d \over {dx}}\right)^n (e^{-x}x^{n+\alpha})$
is then evident again for $h=0$.  The Laguerre polynomials appear, for instance, in the
wave function for the hydrogen atom.

\bigskip
\centerline{\bf Generalized raising and lowering operators for the Legendre equations}
\bigskip
We now focus on the Legendre differential equation ($x=\cos \theta$, $-1 \leq x \leq 1$)
$$\left[{d \over {dx}}(x^2-1){d \over {dx}}-n(n+1)\right]P_n(x)=0, \eqno(2.1)$$
and 
the associated Legendre differential equation
$$\left[{d \over {dx}}(x^2-1){d \over {dx}}+{m^2 \over {1-x^2}}-n(n+1)\right]P_n^m(x)=0, \eqno(2.2)$$
with $|m|<n$.  The associated Legendre functions are related to the Legendre polynomials via
$$P_n^m(x)=(1-x^2)^{m/2}{{d^m P_n^m(x)} \over {dx^m}}.$$
Note that for $m$ odd, $P_n^m(x)$ is algebraic, not polynomial, in $x$.

Taking the raising operator for (2.1) to be
$$R_n=(x^2-1){d \over {dx}}+nx=f_1 {d \over {dx}}g_2+h, \eqno(2.3)$$
we find, multiplicative constants henceforth being omitted,
$$g_2(x)=(x^2-1)^{n/2}\exp\left[-\int {{h(x)} \over {x^2-1}}dx\right],$$
and
$$f_1(x)={{x^2-1} \over {g_2(x)}}=(x^2-1)^{1-n/2}\exp\left[\int {{h(x)} \over {x^2-1}}dx\right].$$
As a brief example, when $h(x)=ax^2+bx+c$, so that
$$\int {{ax^2+bx+c} \over {x^2-1}}dx=ax+{1 \over 2}[(a+b+c)\ln(1-x)+(b-a-c)\ln(1+x),$$
the exponential factor becomes
$$\exp\left[\int {{h(x)} \over {x^2-1}}dx\right]=e^{ax}(1-x)^{(a+b+c)/2}(1+x)^{(b-a-c)/2}.$$
Proceeding similarly with the lowering operator for (2.1),
$$L_n=-{d \over {dx}}(x^2-1)+nx=-g_2 {d \over {dx}}f_1+h,$$
we determine that
$$f_1(x)=(x^2-1)^{1-n/2}\exp\left[\int {{h(x)} \over {x^2-1}}dx\right],$$
and
$$g_2(x)={{x^2-1} \over {f_1(x)}}=(x^2-1)^{n/2}\exp\left[-\int {{h(x)} \over {x^2-1}}dx\right].$$

Generalizing the $m$-raising operator for the associated Legendre equation (2.2),
$$R_m=(1-x^2)^{1/2}{d \over {dx}}+{{mx} \over {(1-x^2)^{1/2}}}=f_1{d \over {dx}}g_2+h,$$
we have $f_1(x)=(1-x^2)^{1/2}/g_2(x)$ and
$$g_2(x)=(1-x^2)^{-m/2}\exp\left[-\int {{h(x)} \over {(1-x^2)^{1/2}}}dx\right].$$
Similarly for the $m$-lowering operator
$$L_m=-{d \over {dx}}(1-x^2)^{1/2}+{{mx} \over {(1-x^2)^{1/2}}}=-g_2{d \over {dx}}f_1+h,$$
we determine that
$$f_1(x)=(1-x^2)^{(m+1)/2}\exp\left[\int {{h(x)} \over {(1-x^2)^{1/2}}}dx\right],$$
with $g_2(x)=(1-x^2)^{1/2}/f_1(x)$.

These operators satisfy $R_nP_{n-1}(x)=nP_n(x)$, $L_{n+2}P_n(x)=nP_{n-1}(x)$, $R_mP_n^m(x)=P_n^{m+1}(x)$, and $L_mP_n^{m+1}(x)=[n(n+1)-m(m+1)]P_n^m(x)=(n-m)(m+n+1)P_n^m(x)$.
For $h=0$ these results reduce to the raising and lowering operators of (10) and (27) of
\cite{bdas}.

From the generalized raising operator (2.3) we find that
$$P_n(x)={{(x^2-1)^{-n/2+1}} \over {2^{n-1}n!}}\exp\left[\int {{h(x)} \over {x^2-1}}\right]
{d \over {dx}}(x^2-1)^{n/2}\exp\left[-\int {{h(x)} \over {x^2-1}}\right]{d^{n-1} \over {dx^{n-1}}}
(x^2-1)^{n-1}$$
$$+h(x) {d^{n-1} \over {dx^{n-1}}}(x^2-1)^{n-1}. \eqno(2.4)$$
This leads to the equation $2^{n-1}n!P_n(x)=R_n2^{n-1}(n-1)!P_{n-1}(x)$.
The relation (2.4) is a generalization of (16) of \cite{bdas} for nonzero functions $h(x)$.

We let
$$e(x) \equiv \exp\left[\int {{h(x)} \over {x^2-1}}\right]. \eqno(2.5)$$
Then we may arrive at families of generalized Rodrigues' formulae in the form
$$n!P_n(x)=(x^2-1)^{1-n/2}[e(x)]^{n-1} {d \over {dx}}\left[(x^2-1)^{3/2}{d \over {dx}}\right]^{n-1} {{(x^2-1)^{1/2}} \over {e(x)}}+F[h],$$
wherein $F$, being $0$ for $h=0$, depends upon powers of $h$ and its derivatives with
respect to $x$.  The term of $F$ with the highest order derivative of $h$ is
$x(x^2-1)^{n-2}h^{(n-2)}(x)$ and the term of the highest power of $h$ is $(-1)^{n}(n-1)!xh^{n-1}(x)$.  

\bigskip
{\it Generalization to Gegenbauer polynomials}

One of the generalizations of Legendre polynomials is the Gegenbauer (or ultraspherical) 
polynomials $C_n^\lambda$, in turn a special case of the Jacobi polynomials 
$P_n^{(\alpha,\beta)}$, \cite{andrews,grad}
$$C_n^\lambda(x)={{(2\lambda)_n} \over {(\lambda+1/2)_n}}P_n^{(\lambda-1/2,\lambda-1/2)}(x).$$
These polynomials satisfy the ordinary differential equation
$$(x^2-1)y''+(2\lambda+1)xy'-n(2\lambda+n)y=0,$$  
and the case $\lambda=1/2$ recovers the Legendre polynomials.  A pair of lowering and
raising operators in $n$ for the Gegenbauer polynomials is given by
$$\left[(1-x^2){d \over {dx}}+nx\right]C_n^\lambda(x)=(n+2\lambda-1)C_{n-1}^\lambda(x),$$
$$\left[(1-x^2){d \over {dx}}-(n-1+2\lambda)x\right]C_{n-1}^\lambda(x)=-nC_n^\lambda(x).$$

Putting
$$C_n^+=(1-x^2){d \over {dx}}-(n-1+2\lambda)x=f_1{d \over {dx}}g_2+h,$$
we find that
$$g_2(x)=(1-x^2)^{(n-1+2\lambda)/2}\exp\left[\int {{h(x)} \over {x^2-1}}dx\right],$$
and
$$f_1(x)={{1-x^2} \over {g_2(x)}}=(1-x^2)^{(3-n)/2-\lambda}\exp\left[-\int {{h(x)} \over {x^2-1}}dx\right].$$
With $C_0^\lambda(x)=1$, via iteration of $C_n^\lambda(x)=-{1 \over n}C_n^+C_{n-1}^\lambda(x)$
for $h=0$ we determine that
$$C_n^\lambda(x)={{(-1)^n} \over {n!}}(1-x^2)^{(3-n)/2-\lambda}{d \over {dx}}\left[(1-x^2)^{3/2}
{d \over {dx}}\right]^{n-1}(1-x^2)^\lambda.$$
For comparison, the usual Rodrigues' formula for Gegenbauer polynomials is
$$C_n^\lambda(x)={{(-2)^n} \over {n!}}{{\Gamma(n+\lambda)\Gamma(n+2\lambda)} \over {\Gamma(
\lambda)\Gamma(2n+2\lambda)}}(1-x^2)^{1/2-\lambda}{d^n \over {dx^n}}(1-x^2)^{n+\lambda-1/2}.$$
From the latter formula we determine that 
$$C_n^\lambda(x)=-{{(-2)^{n-1}} \over {n!}}{{\Gamma(n+\lambda-1)\Gamma(n+2\lambda-1)} \over {\Gamma
(\lambda)\Gamma(2n+2\lambda-2)}}$$
$$\times (1-x^2)^{(3-n)/2-\lambda}{d \over {dx}}(1-x^2)^{n/2}{d^{n-1} \over
{dx^{n-1}}} (1-x^2)^{n+\lambda-3/2}.$$  


\bigskip
\centerline{\bf A Legendre polynomial identity}
\bigskip

The following identity, useful for deriving a Rodrigues' type formula, was given without proof 
in \cite{bdas}.  We present three proofs of this identity, the first using correspondences
with the Legendre polynomials, and the other two employing the product (Leibniz) rule.
For $n \geq 2$,
$$(x^2-1){d^n \over {dx^n}}(x^2-1)^{n-1}=(n-1)n{d^{n-2} \over {dx^{n-2}}}(x^2-1)^{n-1}. \eqno(3.1)$$

For the first proof, we begin by making various identifications of the left side of the identity (3.1) in terms of Legendre polynomials $P_n(x)$ and associated Legendre functions $P_n^m(x)$ \cite{nbs,boas,grad}.  From the usual Rodrigues' formula 
$$P_\ell(x)={1 \over {2^\ell \ell!}}{d^\ell \over {dx^\ell}}(x^2-1)^\ell, \eqno(3.2)$$
the left side of (3.1) is
$$2^{n-1}(n-1)!(x^2-1){d \over {dx}}P_{n-1}(x)=-2^{n-1}(n-1)!(1-x^2)^{1/2}P_{n-1}^1(x),$$
where we have used
$P_\ell^1(x)=(1-x^2)^{1/2}{d \over {dx}}P_\ell(x)$.
From recurrence 12.5.8(f) of \cite{boas} we also have
$$2^{n-1}(n-1)!(x^2-1){d \over {dx}}P_{n-1}(x)=2^{n-1}n![P_n(x)-xP_{n-1}(x)]. \eqno(3.3)$$

The right side of (3.1) is the integral of the polynomial
$$(n-1)n{d^{n-1} \over {dx^{n-1}}}(x^2-1)^{n-1} = 2^{n-1}(n-1)n!P_{n-1}(x).$$
By integrating the recurrence 12.5.8(b) of \cite{boas} by parts we obtain
$$(n+1)\int P_n(x)dx=-P_{n-1}(x)+xP_n(x), \eqno(3.4)$$
and
$$(n+1)\int_0^x P_n(x)dx=-P_{n-1}(x)+xP_n(x)+P_{n-1}(0).$$
The right side of (3.1) is, using (3.4),
$$(n-1)n\int{d^{n-1} \over {dx^{n-1}}}(x^2-1)^{n-1}dx=2^{n-1}(n-1)n!\int P_{n-1}(x)dx$$
$$=(n-1)!(n-1)2^{n-1}[-P_{n-2}(x)+xP_{n-1}(x)]$$
$$=n!2^{n-1}[P_n(x)-xP_{n-1}(x)]. \eqno(3.5)$$
In the last step we have used recurrence 12.5.8(a) of \cite{boas} in order to eliminate the $P_{n-2}(x)$ term.  The equality of (3.3) and (3.5) verifies identity (3.1). \qed

{\it Remark}.  The expressions (3.3) or (3.5) verify that (3.1) evaluates to $0$ when $x=\pm 1$.

A second proof of (3.1) may be based upon the product rule and the factorization
$x^2-1=(x-1)(x+1)$.  We then obtain for the left side of (3.1)
$$(x^2-1){d^n \over {dx^n}}(x^2-1)^{n-1}=(-1)^n \sum_{\ell=0}^n {n \choose \ell}(1-n)_\ell(1-n)_{n-\ell}(x-1)^\ell (x+1)^{n-\ell}, \eqno(3.6)$$
while the right side is given by
$$n(n-1)(-1)^n \sum_{\ell=0}^{n-2}{{n-2} \choose \ell}(1-n)_{n-\ell-2}(1-n)_\ell (x-1)^{\ell+1}
(x+1)^{n-\ell-1}$$
$$=n(n-1)(-1)^n \sum_{\ell=1}^{n-1}{{n-2} \choose {\ell-1}}(1-n)_{n-\ell-1}(1-n)_{\ell-1} (x-1)^\ell (x+1)^{n-\ell}. \eqno(3.7)$$
Since in (3.6), $(1-n)_n=0$ unless $n=0$, the $\ell=0$ and $\ell=n$ terms evaluate to $0$.
Comparing (3.6) and (3.7) we need to verify the equality
$$n(n-1){{n-2} \choose {\ell-1}}(1-n)_{n-\ell-1}(1-n)_{\ell-1}
={n \choose \ell}(1-n)_\ell(1-n)_{n-\ell}.$$
With the ratios $(1-n)_{n-\ell}/(1-n)_{n-\ell-1}=-\ell$ and $(1-n)_{\ell}/(1-n)_{\ell-1}=\ell-n$,
this equality reduces to the readily confirmed statement that
$$n(n-1){{n-2} \choose {\ell-1}}={n \choose \ell}(n-\ell)\ell={{n!} \over {(\ell-1)!(n-\ell-1)!}}.$$
\qed

For a third proof, we first binomially expand each side of (3.1), and then differentiate
term by term:
$${d^{n-2} \over {dx^{n-2}}}(x^2-1)^{n-1}={d^{n-2} \over {dx^{n-2}}}\sum_{\ell=0}^{n-1}{{n-1}
\choose \ell}(-1)^{n-\ell-1}x^{2\ell}$$
$$=-2(-1)^n\sum_{\ell=0}^{n-1}{{n-1}\choose \ell}(-1)^{n-\ell-1}\ell(1-2\ell)_{n-3}x^{2\ell-n+2},$$
and
$${d^n \over {dx^n}}(x^2-1)^{n-1}={d^n \over {dx^n}}\sum_{k=0}^{n-1}{{n-1}\choose k}(-1)^{n-k-1} x^{2k}$$
$$=-2(-1)^n\sum_{k=0}^{n-1}{{n-1}\choose k}(-1)^{n-k-1}k(1-2k)_{n-1}x^{2k-n}.$$
Therefore, the left side of (3.1) is given by
$$-2(-1)^n\sum_{k=0}^{n-1}{{n-1}\choose k}(-1)^{n-k-1}k(1-2k)_{n-1}(x^{2k-n+2}-x^{2k-n})$$
$$=-2(-1)^n\left[\sum_{k=0}^{n-1}{{n-1}\choose k}(-1)^{n-k-1}k(1-2k)_{n-1}x^{2k-n+2}\right.$$
$$\left.-\sum_{k=0}^{n-2}{{n-1}\choose {k+1}}(-1)^{n-k}(k+1)(-1-2k)_{n-1}x^{2k-n+2}\right].$$
Thus, equating the coefficients of $x^{2k-n+2}$, (3.1) is valid if the following equality is verified:
$${{n-1} \choose k}(-1)^{n-k-1}k(1-2k)_{n-1}-{{n-1} \choose {k+1}}(-1)^{n-k}(k+1)(-1-2k)_{n-1}$$
$$=n(n-1){{n-1} \choose k}(-1)^{n-k-1}k(1-2k)_{n-3}. \eqno(3.8)$$
We have the ratios
$${{(1-2k)_{n-3}} \over {(-1-2k)_{n-1}}}={1 \over {2k(2k+1)}}, ~~~~~~
{{(1-2k)_{n-1}} \over {(1-2k)_{n-3}}}=(2k-n+1)(2k-n+2),$$
and
$${{(1-2k)_{n-1}} \over {(-1-2k)_{n-1}}}={{(2k-n+1)(2k-n+2)} \over {2k(2k+1)}}, ~~~~~~
{{{n-1} \choose {k+1}} \over {{n-1} \choose k}}={{n-k-1} \over {k+1}}.$$
Then indeed (3.8) reduces to the equality
$${{(2k-n+1)(2k-n+2)} \over {2k+1}}+2(n-k-1)={{n(n-1)} \over {2k+1}}.$$
\qed

{\it Remark}.  It may be shown that (3.1) is equivalent to the three-term identity
$$(x^2-1){d^{n-1} \over {dx^{n-1}}}(x^2-1)^{n-1}-2x{d^{n-2} \over {dx^{n-2}}}(x^2-1)^{n-1}
=(n+1)(n-2){d^{n-3} \over {dx^{n-3}}}(x^2-1)^{n-1},$$
which in terms of Legendre polynomials is the equality
$$(x^2-1)P_{n-1}(x)+{{2x} \over {2n-1}}[P_{n-2}(x)-P_n(x)]$$
$$={{(n+1)(n-2)} \over {1-2n}}\left\{{1 \over {2n-3}}[P_{n-1}(x)-P_{n-3}(x)]+
{1 \over {2n+1}}[P_{n-1}(x)-P_{n+1}(x)]\right\}.$$


\bigskip
\centerline{\bf Discussion}
\bigskip

For the Chebyshev polynomials of the second kind $U_n(x)$ \cite{mason,rivlin},
a set of lowering and raising operators is given by
$$\left[(1-x^2){d \over {dx}}+nx\right]U_n(x)=(n+1)U_{n-1}(x),$$
and
$$\left[(x^2-1){d \over {dx}}+(n+2)x\right]U_n(x)=(n+1)U_{n+1}(x).$$
With $U^+_n=(x^2-1){d \over {dx}}+(n+2)x=f_1 {d \over {dx}}g_2+h$, we may replace
$n \to n+2$ in the result of (2.3), giving
$$g_2(x)=(x^2-1)^{n/2+1}{1 \over {e(x)}},$$
and
$$f_1(x)={{x^2-1} \over {g_2(x)}}=(x^2-1)^{-n/2}e(x),$$
with $e(x)$ as given in (2.5).  Taking the special case of $h=0$, with $U_0=1$ and
the iteration $n!U_n(x)=U^+_{n-1}U^+_{n-2}\cdots U^+_1U^+_0 U_0$,
we obtain
$$n!U_n(x)=(x^2-1)^{(1-n)/2}{d \over {dx}}\left[(x^2-1)^{3/2}{d \over {dx}}\right]^{n-1}(x^2-1).$$
This expression may be contrasted with the trigonometric definition 
$$U_n(x)={{\sin[(n+1)\cos^{-1} x]} \over {\sin(\cos^{-1} x)}},$$
and the usual Rodrigues' formula
$$U_n(x)={{(-1)^n(n+1)} \over {\sqrt{1-x^2}(2n+1)!!}}{d^n \over {dx^n}}(1-x^2)^{n+1/2},$$
where $(2n+1)!!=(2n+1)(2n-1)\cdots 5 \cdot 3 \cdot 1$.  Using this formula and
$U_n(x)={1 \over n}U_{n-1}^+U_{n-1}(x)$ gives the expression
$$U_n(x)={{(-1)^n} \over {(2n-1)!!}}(1-x^2)^{(1-n)/2}{d \over {dx}}(1-x^2)^{n/2}{d^{n-1} \over
{dx^{n-1}}} (1-x^2)^{n-1/2}.$$

The Chebyshev polynomials of the first kind $T_n(x)$ satisfy the differential equation
$(1-x^2)u''-xu'+n^2u=0$ and may be defined from $T_n(x)=\cos[n \cos^{-1}x]$.  Raising and
lowering operators $T^\pm_m$ such that $T^\pm_m T_m(x)=T_{m \pm 1}(x)$, are given by, for
$m>0$,
$$T_m^\pm = \mp {{(1-x^2)} \over m}{d \over {dx}}+x.$$
Putting $T^+_m=f_1 {d \over {dx}}g_2+h$, we find that
$$g_2(x)=(1-x^2)^{m/2}\exp\left[m \int {{h(x)} \over {1-x^2}}dx\right],$$
and 
$$f_1(x)=-{{(1-x^2)} \over {mg_2(x)}}=-{1 \over m}(1-x^2)^{-m/2+1}\exp\left[-m \int {{h(x)} \over {1-x^2}}dx\right].$$
We have $T_0=1$, $T_1(x)=x$, and
$T_m(x)=T^+_{m-1}T^+_{m-2}\cdots T^+_1 T_1(x)$,
giving  for $h=0$
$$T_n(x)={{(-1)^{n-1}}\over {(n-1)!}}(1-x^2)^{(3-n)/2}{d \over {dx}}\left[(1-x^2)^{3/2}
{d \over {dx}}\right]^{n-2}x(1-x^2)^{1/2}.$$
In comparison, the usual Rodrigues' formula for $T_n(x)$ is
$$T_n(x)={{(-1)^n \sqrt{1-x^2}} \over {(2n-1)!!}}{d^n \over {dx^n}}(1-x^2)^{n-1/2}.$$
Using this formula and
$T_n(x)=T_{n-1}^+T_{n-1}(x)$ gives the expression
$$T_n(x)={{(-1)^n} \over {(2n-3)!!(n-1)}}(1-x^2)^{(3-n)/2}{d \over {dx}}(1-x^2)^{n/2}{d^{n-1} \over
{dx^{n-1}}} (1-x^2)^{n-3/2}.$$

With a change of variable $x=\cos y$, the raising and lowering operators for $T_n$ and $U_n$
have been identified as elements of a $su(1,1)$ algebra for the infinitely deep square well
potential.  They were also used to study the temporally stable coherent states for the
infinite well and P\"{o}schl-Teller potentials \cite{antoine}.  

For the radial Coulomb problem, a raising operator may take the form
$$A_\ell^+ ={d \over {dr}}+{{\ell+1} \over r}=f_1(r){d \over {dr}}g_2(r)+h(r),$$
with angular momentum quantum number $\ell$ (cf. \cite{coffey,ding,fernandez}).
Then we find $f_1(r)=1/g_2(r)$ and
$$g_2(r)=r^{\ell+1} \exp\left[-\int h(r) dr\right].$$
Similarly, for a three dimensional harmonic oscillator with raising operator
$$a_\ell^+ ={d \over {dr}}+{r \over 2}+{{\ell+1} \over r}=f_1(r){d \over {dr}}g_2(r)+h(r),$$
we have $f_1(r)=1/g_2(r)$ and
$$g_2(r)=r^{\ell+1} e^{r^2/4}\exp\left[-\int h(r) dr\right].$$
For these two (scaled) radial problems, the wave function contains products of powers of $r$,
a decreasing exponential factor $\exp(-r/2)$, and Laguerre polynomials.


Many more applications of our approach are possible.  In some more generality,
suppose, for example, that a lowering operator is given as
$$L=-{d \over {dx}}a(x)+b(x)=-g_2{d \over {dx}}f_1+h,$$
wherein $a$ and $b$ are specified functions.  Then we determine that
$$f_1(x)=a(x)\exp\left[-\int {{b(x)} \over {a(x)}}dx\right]\exp\left[\int {{h(x)} \over {a(x)}}dx\right],$$
and
$$g_2(x)={{a(x)} \over {f_1(x)}}=\exp\left[\int {{b(x)} \over {a(x)}}dx\right]\exp\left[-\int {{h(x)} \over {a(x)}}dx\right].$$
Indeed, we have produced generalized lowering and raising operators and a generalized
Rodrigues' formula for arbitrary Jacobi polynomials $P_n^{(\alpha,\beta)}(x)$, and this 
will be reported elsewhere.

Another direction for obtaining Rodrigues-type formulae proceeds as follows, illustrated for
the Laguerre polynomials of the Introduction.  Making the change of variable $x=r^2$, the
differential equation (1.2) becomes 
$${d^2 \over {dr^2}}y(r^2)+{2 \over r}\left(\alpha-r^2+{1 \over 2}\right){d \over {dr}}y(r^2)
+4ny(r^2)=0,$$
and the raising and lowering operators (1.3) transform to
$$A_+={r \over 2}{d \over {dr}}+n+\alpha-r^2,~~~~~~~~
A_-=-{r \over 2}{d \over {dr}}+n.$$
The generalized raising operator $A_+=f_1(r){d \over {dr}}g_2(r)+h(r)$ has
$$g_2(r)=\exp\left[-2\int {{h(r)} \over r}dr\right]e^{-r^2}r^{2(n+\alpha)}$$
and $f_1(r)=r/[2g_2(r)]$.
By iteration with $h=0$ and $L_0^\alpha(r^2)=1$ we obtain
$$L_n^\alpha(r^2)={1 \over {n!}}{r \over 2^n}e^{r^2}r^{-2(n+\alpha)}{d \over {dr}}
\left(r^3{d \over {dr}}\right)^{n-1} e^{-r^2} r^{2(\alpha+1)}.$$
Such relations may be all the more important because of the connection (1.1) to Hermite
polynomials when $\alpha=\pm 1/2$, so that
$$H_{2n}(r)=(-1)^n 2^n r^{-2n}e^{r^2}{d \over {dr}}
\left(r^3{d \over {dr}}\right)^{n-1} e^{-r^2} r,$$
and
$$H_{2n+1}(r)=(-1)^n 2^{n+1}r^{-2n+1}e^{r^2}{d \over {dr}}
\left(r^3{d \over {dr}}\right)^{n-1} e^{-r^2} r^3.$$

In conclusion, we have generalized raising and lowering operators that are so useful for
quantum mechanics.  These then lead to generalized Rodrigues' formulae and other identities.
The generalized operators have been illustrated for the Chebyshev, Gegenbauer, Hermite, Laguerre, and Legendre polynomials.  Especially the latter three polynomials appear in important quantum
mechanical and other problems, including the harmonic oscillator and Coulomb problem.  
Moreover, when combined with Gaussian factors, Hermite and Laguerre polynomials are very useful 
for describing optical fields.
For a variety of problems with spherical symmetry, the Legendre polynomials and associated
Legendre functions appear.  We have provided discussion specific to the latter functions 
that generalize previous results.  Given the existence of generalized Rodrigues' formulae,
there is also the possibility to establish generalized Schlaefli integrals.  The latter
contour integrals also play a role in the theory of orthogonal polynomials.


  
\pagebreak

\end{document}